\documentclass[12pt]{article}
\usepackage{epsfig}
\usepackage{amsfonts}
\usepackage{amscd}
\usepackage{latexsym}
\usepackage{amsmath,amssymb}
\usepackage{verbatim}
\usepackage{setspace}
\usepackage{color}
\usepackage{cite}

\usepackage[textheight=9in, textwidth=6.5in, letterpaper]{geometry}

\usepackage{hyperref}
\hypersetup{
    colorlinks,
    citecolor=black,
    filecolor=black,
    linkcolor=black,
    urlcolor=black
    linktoc=all
}

\numberwithin{equation}{section}

   \makeatletter
  \let\over=\@@over \let\overwithdelims=\@@overwithdelims
  \let\atop=\@@atop \let\atopwithdelims=\@@atopwithdelims
  \let\above=\@@above \let\abovewithdelims=\@@abovewithdelims
\renewcommand\section{\@startsection {section}{1}{\z@}%
                                   {-3.5ex \@plus -1ex \@minus -.2ex}
                                   {2.3ex \@plus.2ex}%
                                   {\normalfont\large\bfseries}}

\renewcommand\subsection{\@startsection{subsection}{2}{\z@}%
                                     {-3.25ex\@plus -1ex \@minus -.2ex}%
                                     {1.5ex \@plus .2ex}%
                                     {\normalfont\bfseries}}

\begin{document}
\unitlength = 1mm

\ \\
\vskip 1cm
\begin{center}

{\textbf{\large{Conformal mapping of the Misner-Sharp mass from gravitational collapse}}}

\vspace{0.8cm}
Fay\c{c}al Hammad\footnote{fhammad@ubishops.ca}

\vspace{1cm}

{\it  Physics Department \& STAR Research Cluster, Bishop's University\\
2600 College Street, Sherbrooke, (QC) J1M 1Z7, Canada\\
Physics Department, Champlain College-Lennoxville\\
2580 College Street, Sherbrooke, (QC) J1M 0C8, Canada}\\

\begin{abstract}
The conformal transformation of the Misner-Sharp mass is reexamined. It has recently been found that this mass does not transform like usual masses do under conformal mappings of spacetime. We show that when it comes to conformal transformations, the widely used geometric definition of the Misner-Sharp mass is fundamentally different from the original conception of the latter. Indeed, when working within the full hydrodynamic setup that gave rise to that mass, i.e. the physics of gravitational collapse, the familiar conformal transformation of a usual mass is recovered. The case of scalar-tensor theories of gravity is also examined.
\end{abstract}
\end{center}

\\

\quad PACS number(s): {97.60.-s, 47.75.+f, 04.40.Dg}
\vspace{1cm}




\def\vx{{\vec x}}
\def\p{\partial}
\def\po{$\cal P_O$}

\pagenumbering{arabic}

\section{Introduction}
Due to the equivalence principle of general relativity, gravity may locally be completely canceled, whence the difficulty of defining a gravitational mass/energy at each spacetime point. As a consequence, only non-local definitions are constructed, i.e. to any given point of spacetime, one only associates a non-vanishing mass/energy to a finite neighborhood of that point.

One of the well-known definitions of a gravitational mass is the so-called Misner-Sharp mass \cite{MS}. This mass is defined for any spherically symmetric spacetime region. The definition has been applied to a large range of physical problems; from the physics of gravitational collapse to the thermodynamics of horizons (see e.g. Refs.~\cite{Hayward1, Nelson, Faraoni1} and the references therein) and cosmology (see also Ref.~\cite{Faraoni1} and the references therein). In the case of a spherically symmetric gravitational collapse, the Misner-Sharp mass represents, at a given radial coordinate, the total energy inside the collapsing spherical shell of matter. In cosmology, the Misner-Sharp mass represents, at a given radial coordinate, the total energy, enclosed inside the spherical shell, of the cosmic fluid of the Universe.

The practicality of the Misner-Sharp mass stems from its simple geometric definition. For a spherically symmetric spacetime region, whose metric is $g_{\mu\nu}$ and whose areal radius is $R$, the corresponding Misner-Sharp mass $m(r,t)$ is such that $1-2Gm(r,t)/R=g^{\mu\nu}\partial_{\mu}R\partial_{\nu}R$, where $G$ is Newton's gravitational constant, and $r$ and $t$ are the co-moving radial and time coordinates, respectively. Based on this definition, the Misner-Sharp mass looks more like a purely geometric quantity than a material entity. As such, the mass would certainly behave under spacetime transformations, conformal ones in particular, just as geometry would. In contrast to geometry, however, matter content behaves differently. This fact is indeed the cause behind the well-known non-invariance of Einstein equations under conformal transformations (see e.g. Refs.~\cite{Fujii,Faraoni2,Capozzieollo}.) As we shall see in this paper, this fact turns out also to be the cause behind the 'wrong' conformal transformation of the Misner-Sharp mass found recently in Ref.~\cite{Faraoni3}.

The main issue of the conformal transformations, is not so much the fact that additional terms appear on the right-hand side of the transformed Einstein equations, as these are easily interpreted as representing the 'work' done to curve space-time during the conformal transformation. The transformed field equations are viewed then as acquiring these additional terms because of the induced energy-momentum tensor corresponding to this 'work'. The real issue is actually that the old terms themselves do not conserve their forms. The conformal factor multiplying the matter content on the right-hand side of the field equations is different from that multiplying the geometry on the left-hand side.

Now, this second issue does find a solution in scalar-tensor theories of gravity, like Brans-Dicke theory \cite{BD}. This is indeed solved thanks to the fact that in these theories, Newton's gravitational constant $G(\phi)$ is taken as depending on the Brans-Dicke scalar field $\phi$ and, hence, is also affected by conformal transformations. The combined transformation in Brans-Dicke theory of $G(\phi)=1/\phi$ and matter on the right-hand side of Einstein equations just matches the transformation of the purely geometric left-hand side of the equations.

From the purely geometric definition above, one might expect then that the Misner-Sharp mass will behave in Brans-Dicke theory just like the field equations, i.e. that after a conformal transformation, the new mass will acquire the right conformal factor as the one acquired by matter on the right-hand side, recovering the way usual masses transform under conformal mappings of spacetime. This would, however, suggest that under conformal transformations the Misner-Sharp mass behaves as a 'normal' mass in scalar-tensor theories of gravity but behaves as an 'exotic' mass in general relativity. As we shall see in this paper, when using the original hydrodynamic definition of the Misner-Sharp mass, introduced in Ref.~\cite{MS}, one recovers the right conformal transform in both frameworks.

The outline of the rest of this paper is as follows. In Sec.~\ref{sec:2}, we briefly recall the general theory of conformal transformations of Einstein field equations and give the interpretation of the result found in Ref.~\cite{Faraoni3} for the conformal transformation of the Misner-Sharp mass based on its usual geometric definition. The result is then compared to what is found in the framework of the Brans-Dicke theory. In Sec.~\ref{sec:3}, we recall the hydrodynamic setup behind the geometric definition of the Misner-Sharp mass and show how the definition is altered in the conformal frame. We end this paper with a brief conclusion and discussion section.

\section{Conformal Mappings of Einstein Equations and of the Geometric Definition of Misner-Sharp Mass}\label{sec:2}
In this section, we recall the main results that we will use later when we refer to the conformally transformed Einstein field equations (see e.g. Refs.~\cite{Fujii,Faraoni2,Capozzieollo} for details). We also review the effect of the conformal transformations on the Misner-Sharp mass when the purely geometric definition of the latter is adopted \cite{Faraoni3}.

A conformal transformation is that transformation of spacetime which changes the metric from $g_{\mu\nu}$ to a new metric $\tilde{g}_{\mu\nu}$ thanks to a non-vanishing spacetime-dependent conformal factor, usually denoted $\Omega(x)$, such that,
\begin{equation}\label{ConfMetric}
\tilde{g}_{\mu\nu}=\Omega^{2}(x)g_{\mu\nu}.
\end{equation}
The Hilbert-Einstein action $\int\mathrm{d}^{4}x\sqrt{-g}(\mathcal{R}+\mathcal{L}_{m})$, where $\mathcal{R}$ is the Ricci scalar and $\mathcal{L}_{m}$ is the matter Lagrangian, does not remain invariant after the transformation (\ref{ConfMetric}) of the metric, nor does the set of Einstein field equations,
$G_{\mu\nu}\equiv\mathcal{R}_{\mu\nu}-\frac{1}{2}g_{\mu\nu}\mathcal{R}=\kappa T_{\mu\nu}$. In what follows, $T_{\mu\nu}$ will denote the energy-momentum tensor of matter and Newton's gravitational constant $G$ will appear inside $\kappa=8\pi G$.

Under the conformal transformation (\ref{ConfMetric}), the Einstein tensor $G_{\mu\nu}$ transforms into $\tilde{G}_{\mu\nu}$ and the energy-momentum tensor transforms into $\tilde{T}_{\mu\nu}=\Omega^{-2}T_{\mu\nu}$. However, the conformally transformed Einstein equations will not read $\tilde{G}_{\mu\nu}=\kappa \tilde{T}_{\mu\nu}$, but will take instead the following form:
\begin{equation}\label{ConfFieldEq}
\tilde{G}_{\mu\nu}=\kappa T_{\mu\nu}+T_{\mu\nu}^{\Omega},
\end{equation}
where $T_{\mu\nu}^{\Omega}$ is introduced for convenience to denote a set of first- and second-order derivatives of the conformal factor $\Omega$. It can be interpreted as an induced energy-momentum tensor, associated with the 'work' done during the conformal transformation to deform the metric, i.e., to deform spacetime \cite{Dabrowski}. Its explicit expression reads \cite{Dabrowski},
\begin{equation}\label{InducedT}
T_{\mu\nu}^{\Omega}=\frac{4\nabla_{\mu}\Omega\nabla_{\nu}\Omega}{\Omega^{2}}-\frac{2\nabla_{\mu}\nabla_{\nu}\Omega}{\Omega}
+g_{\mu\nu}\left(2\frac{\square\Omega}{\Omega}-\frac{\nabla_{\rho}\Omega\nabla^{\rho}\Omega}{\Omega^{2}}\right).
\end{equation}

In view of our later needs, we include here a simple application of the above formulae. Let us perform a conformal transformation with conformal factor $\Omega(t)$ on the spatially flat and homogeneous Friedmann-Lema\^{\i}tre-Robertson-Walker (FLRW) Universe, filled with a perfect fluid and whose metric in the coordinates $(t,r,\theta,\varphi)$ reads,
\begin{equation}\label{FLRW}
\mathrm{d}s^{2}=-\mathrm{d}t^{2}+a^{2}(t)(\mathrm{d}r^{2}+r^{2}\mathrm{d}o^{2}),
\end{equation}
where $a(t)$ is the positive scale factor and $\mathrm{d}o^{2}=\mathrm{d}\theta^{2}+\sin^{2}\theta\mathrm{d}\varphi^{2}$ is the line element on the unit two-sphere.

The Friedmann equation $H^{2}=\frac{1}{3}\kappa\rho$ is obtained from the time-time component of the field equations, $G_{t}^{\,t}=\kappa T_{t}^{\,t}$ after using the fact that $-T_{t}^{\,t}$ is just the mass density $\rho$ of the perfect fluid and that $G_{t}^{\,t}=-3H^{2}$. The Hubble parameter being $H=\dot{a}/a$, with the overdot denoting a derivative with respect to the co-moving time $t$. After a conformal transformation, the new Friedmann equation will not take the simple form $\tilde{H}^{2}=\frac{1}{3}\kappa\tilde{\rho}$, since the form of Einstein equations are not preserved. Instead, the new Friedmann equation, as derived from the conformally transformed field equations (\ref{ConfFieldEq}) by taking the time-time component $\tilde{G}_{t}^{\,t}=\Omega^{-2}(\kappa T_{t}^{\,t}+T^{\Omega t}_{\;\,t})$ and using (\ref{InducedT}), reads,
\begin{equation}\label{ConfFriedmann}
\tilde{H}^{2}=\frac{1}{\Omega^{2}}\left(\frac{\kappa\rho}{3}+2H\frac{\dot{\Omega}}{\Omega}+\frac{\dot{\Omega}^{2}}{\Omega^{2}}\right).
\end{equation}
The last two terms in this formula are interpreted as due to the energy density needed to be supplied to deform spacetime \cite{Dabrowski}. This interpretation becomes more evident when (\ref{ConfFriedmann}) is used for the case of a conformal transformation (\ref{ConfMetric}) with $\Omega=a(t)$, that brings the flat Minkowski spacetime, $\mathrm{d}s^{2}=-\mathrm{d}t^{2}+\mathrm{d}r^{2}+r^{2}\mathrm{d}o^{2}$, to the conformally flat FLRW spacetime, $\mathrm{d}\tilde{s}^{2}=-\mathrm{d}\eta^{2}+a^{2}(\mathrm{d}r^{2}+r^{2}\mathrm{d}o^{2})$, where $\eta$ is the conformal time such that $\mathrm{d}\eta=a\mathrm{d}t$. Indeed, in this case, the original density $\rho$ and Hubble parameter $H$ both vanish and (\ref{ConfFriedmann}) reduces to $\tilde{H}^{2}\equiv (a_{,\eta})^{2}/a^{2}=\dot{\Omega}^{2}/\Omega^{4}$, with the last term playing the role of $\kappa\rho_{_{\Omega}}/3$ in the Friedmann equation.

Let us now recall the conformal transformation of the geometric definition of the Misner-Sharp mass $m(t,r)$. The definition that is mostly used in the literature is the following:
\begin{equation}\label{MS}
m(t,r)=\frac{R}{2G}\left(1-g^{\mu\nu}\partial_{\mu}R\partial_{\nu}R\right),
\end{equation}
where $R$ is the areal radius of the spherical region of spacetime under study. For the FLRW metric (\ref{FLRW}), the areal radius is just $R(t,r)=a(t)r$. Therefore, the above formula yields \cite{Faraoni3},
\begin{equation}\label{MSinGR}
m(t,r)=\frac{R^{3}}{2G}H^{2}=\frac{4\pi R^{3}}{3}\rho.
\end{equation}
The second equality has been obtained by substituting the Friedmann identity, $H^{2}=\kappa\rho/3$. This result is physically attractive for the fact that the last equality in (\ref{MSinGR}) represents nothing but the total mass of the perfect fluid inclosed by the spherical shell of radius $R$. Thus, the important thing to retain here is that the physical meaning in terms of matter content of the geometric definition of the mass $m(t,r)$ is achieved only after using Einstein equations in the form of the Friedmann identity. Therefore, as long as the form of Einstein equations is preserved, it does not matter which side of these equations is used to define the Misner-Sharp mass. As recalled above, however, under conformal transformations the two sides transform differently and, consequently, different conformal transformations for the mass are obtained depending on which side one is relying on.

Let us then briefly review here the conformal transformation of the mass $m(t,r)$ when defined using pure geometry as in (\ref{MS}).
Performing the transformation (\ref{ConfMetric}) with a conformal factor depending only on time, to preserve the homogeneity of the FLRW metric, the latter transforms into,
\begin{equation}\label{ConfFLRW}
\mathrm{d}\tilde{s}=-\Omega^{2}(t)\mathrm{d}t^{2}+\Omega^{2}(t)a^{2}(t)(\mathrm{d}r^{2}+r^{2}\mathrm{d}o^{2}),
\end{equation}
giving the new areal radius $\tilde{R}(t,r)=\Omega(t)R(t,r)$. On the other hand, relying on identity (\ref{MS}), the conformally transformed Misner-Sharp mass one finds would read, $\tilde{m}(t,r)=\frac{1}{2G}\tilde{R}(1-\tilde{g}^{\mu\nu}\partial_{\mu}\tilde{R}\partial_{\nu}\tilde{R})=\frac{1}{2G}\tilde{R}^{3}\tilde{H}^{2}$. Therefore, given that the Hubble parameter transforms as
\begin{equation}\label{ConfH}
\tilde{H}^{2}=\frac{H^{2}}{\Omega^{2}}+2H\frac{\dot{\Omega}}{\Omega^{3}}+\frac{\dot{\Omega}^{2}}{\Omega^{4}},
\end{equation}
as it follows from a direct computation of $\tilde{H}=\tilde{a}^{-1}\mathrm{d}\tilde{a}/\mathrm{d}\eta$, where $\eta$ is the conformal time defined by $\mathrm{d}\eta=\Omega\mathrm{d}t$, the transformed mass $\tilde{m}(t,r)$ one finds is,
\begin{equation}\label{ConfMSinGR}
\tilde{m}(t,r)=\Omega\left[m(t,r)+\frac{R^{3}}{2G}\left(2H\frac{\dot{\Omega}}{\Omega}+\frac{\dot{\Omega}^{2}}{\Omega^{2}}\right)\right].
\end{equation}
Note that this result is also what is obtained when using the field equations, i.e. the transformed Friedmann equation (\ref{ConfFriedmann}) together with the second equality in (\ref{MSinGR}). We recognize in the last two terms inside the parenthesis the energy density we interpreted above as coming from the induced energy-momentum tensor (\ref{InducedT}). The non-material nature of this induced energy density, exhibiting no interaction between the last term and the original mass $m(t,r)$, might already lead us to suspect that it is just an artifact of the geometric nature of the mass and that one might simply not take it into account in the transformed mass if one is only interested in the fate of the material content. Nevertheless, the interesting thing about (\ref{ConfMSinGR}) is that both the geometrically induced energy density and the original mass $m(t,r)$, appearing in the first term, acquire the same 'wrong' conformal factor. Indeed, a mass/energy density transforms under a conformal transformation as $\tilde{\rho}=\Omega^{-4}\rho$ and a three-volume as $\tilde{\mathrm{V}}=\Omega^{3}\mathrm{V}$, so that a 'normal' mass would transform as $m\rightarrow\tilde{m}=\tilde{\mathrm{V}}\tilde{\rho}=\Omega^{-1}m$.

At this point, one might argue that this issue is only peculiar to the energy density of the Universe since density in an FLRW Universe is bound to geometry through the Hubble parameter. However, the issue becomes actually even worse when dealing with a single black hole. Let us examine a Schwarzschild black hole of constant mass $M$, and whose metric is,
\begin{equation}\label{SchwMetric}
\mathrm{d}s^{2}=-f(r)\mathrm{d}t^{2}+f(r)^{-1}\mathrm{d}r^{2}+r^{2}\mathrm{d}o^{2}.
\end{equation}
Here, $f(r)=1-2GM/r$. Since the areal radius of this black hole is simply $r$, the definition (\ref{MSinGR}) of the Misner-Sharp mass will give for this geometry $m(t,r)=M$. This is a reasonable result since the only physical mass contained inside the spacetime is the mass $M$ at the center of the collapsed matter inside the black hole. Therefore, one might expect that for this case the Misner-Sharp mass would certainly transform as a normal mass does, i.e. that $\tilde{m}(t,r)=M/\Omega$. However, under the conformal transformation (\ref{ConfMetric}), with a conformal factor $\Omega(t,r)$ depending only on the radial and time coordinates to preserve spherical symmetry, the Schwarzschild metric (\ref{SchwMetric}) becomes
\begin{equation}\label{ConfSchwMetric}
\mathrm{d}s^{2}=-\Omega^{2}f(r)\mathrm{d}t^{2}+\Omega^{2}f^{-1}(r)\mathrm{d}r^{2}+\Omega^{2}r^{2}\mathrm{d}o^{2}.
\end{equation}
From this expression, we see that the new areal radius of the black hole is $\tilde{R}=\Omega r$. The new Misner-Sharp mass one finds for this new metric when one relies on the definition (\ref{MS}) and writes $\tilde{m}(t,r)=\frac{1}{2G}\tilde{R}(1-\tilde{g}^{\mu\nu}\partial_{\mu}\tilde{R}\partial_{\nu}\tilde{R})$, is
\begin{equation}\label{ConfSchwMS}
\tilde{m}(t,r)=\Omega\left[M+\frac{r^{3}}{2G}\left(\frac{\dot{\Omega}^{2}}{\Omega^{2}f}-2f\left[\frac{\Omega'}{r\Omega}
+\frac{\Omega'^{2}}{2\Omega^{2}}\right]\right)\right].
\end{equation}
where the prime stands for a derivative with respect to the co-moving coordinate $r$. Thus, not only the mass at the center of the black hole transforms in the wrong way, but also no simple interpretation of the rest of the terms could be found.

Now, from the observation made below the result (\ref{ConfMSinGR}), one hopes that by going to scalar-tensor theories of gravity, where both sides of field equations transform in a similar way, one would either recover the same (and the 'right') conformal factor everywhere in (\ref{ConfMSinGR}), or simply obtain a conformally invariant mass $m(t,r)$. Actually, the former option is satisfied when switching to scalar-tensor theories. Indeed, in a Brans-Dicke Lagrangian, $(16\pi)^{-1}\sqrt{-g}(\phi R-\phi^{-1}\omega(\phi)\partial_{\mu}\phi\partial^{\mu}\phi)+L_{_{matter}}$, where $\phi$ is the Brans-Dicke scalar field, $\omega(\phi)$ is the Brans-Dicke parameter and $\mathcal{L}_{m}$ is the matter fields Lagrangian, one has the following field equations:
\begin{equation}\label{BDFieldEq}
G_{\mu\nu}=\frac{8\pi}{\phi}T_{\mu\nu}+T^{\phi}_{\mu\nu},
\end{equation}
with the energy-momentum tensor $T^{\phi}_{\mu\nu}$, associated to the field $\phi$, given by
\begin{equation}\label{TPhi}
T^{\phi}_{\mu\nu}=\frac{\omega\nabla_{\mu}\phi\nabla_{\nu}\phi}{\phi^{2}}+\frac{\nabla_{\nu}\nabla_{\mu}\phi}{\phi}-g_{\mu\nu}\left(\frac{\square\phi}{\phi}
+\frac{\omega\nabla_{\sigma}\phi\nabla^{\sigma}\phi}{2\phi^{2}}\right)
\end{equation}
From (\ref{BDFieldEq}) and (\ref{TPhi}) one extracts the Friedmann equation by writing explicitly the $tt$-component of the field equations (\ref{BDFieldEq}):
\begin{equation}\label{BDFriedmann}
H^{2}=\frac{8\pi}{3\phi}\rho+\frac{{\omega}}{6}\frac{\dot{\phi^{2}}}{\phi^{2}}-H\frac{\dot{\phi}}{\phi}.
\end{equation}
The field equations (\ref{BDFieldEq}), and hence also the Friedmann equation (\ref{BDFriedmann}), are invariant under the conformal transformation (\ref{ConfMetric}) with a conformal factor $\Omega=\phi^{\alpha}$, provided that the scalar field $\phi$ and the Brans-Dicke parameter $\omega$ transform, respectively, as \cite{Faraoni5},
\begin{equation}\label{BDFactors}
\tilde{\phi}=\Omega^{-2}\phi,\qquad\tilde{\omega}=\frac{\omega-6\alpha(\alpha-1)}{(1-2\alpha)^{2}}.
\end{equation}

Now if one adopts the definition (\ref{MS}) for the Misner-Sharp mass one would use the first equality in (\ref{MSinGR}) but replaces Newton's gravitational constant $G$ there by $1/\phi$, i.e. $m(t,r)=\frac{1}{2}\phi R^{3}{H}^{2}$. Upon substituting the right-hand side of (\ref{BDFriedmann}) for $H^{2}$, one then finds the following explicit expression for $m(t,r)$ in Brans-Dicke theory \cite{Faraoni4}:
\begin{equation}\label{MSinBD}
m(t,r)=\frac{\phi R^{3}}{2}\left(\frac{8\pi}{3\phi}\rho+\frac{\omega}{6}\frac{\dot{\phi}^{2}}{\phi^{2}}-H\frac{\dot{\phi}}{\phi}\right).
\end{equation}
After a conformal transformation, one would write $\tilde{m}(t,r)=\frac{1}{2}\tilde{\phi}\tilde{R}^{3}{\tilde{H}}^{2}$ which gives, after using (\ref{ConfH}), (\ref{BDFriedmann}), and the field redefinition (\ref{BDFactors}), the following expression:
\begin{equation}\label{ConfMSinBD}
\tilde{m}(t,r)=\Omega^{-1}\left[m(t,r)+\frac{\phi R^{3}}{2}\left(2H\frac{\dot{\Omega}}{\Omega}+\frac{\dot{\Omega}^{2}}{\Omega^{2}}\right)\right].
\end{equation}
We notice in this formula the emergence again of the induced energy density and that all the terms acquire the right conformal factor that masses acquire under a conformal transformation. If, instead of the geometric definition (\ref{MS}), one relies on the general result (\ref{MSinGR}) obtained from the field equations, which is here legitimate given that in Brans-Dicke theory the field equations are invariant, one finds
\begin{align}
\tilde{m}(t,r)&=\frac{\tilde{\phi} \tilde{R}^{3}}{2}\left(\frac{8\pi}{3\tilde{\phi}}\tilde{\rho}+\frac{\tilde{\omega}}{6}\frac{\dot{\tilde{\phi}}^{2}}{\tilde{\phi}^{2}}
-\tilde{H}\frac{\dot{\tilde{\phi}}}{\tilde{\phi}}\right),
\end{align}
where the time-derivative should be taken here with respect to the conformal time $\eta$. A straightforward calculation using the transformation properties (\ref{BDFactors}) shows that this new mass takes on exactly the same form as (\ref{ConfMSinBD}). Comparing (\ref{ConfMSinGR}) and (\ref{ConfMSinBD}), leads us to conclude that the Misner-Sharp mass behaves as a 'real' mass under conformal transformation only in conformally invariant scalar-tensor theories of gravity.
\section{Conformal Mapping of a Gravitational Collapse}\label{sec:3}
Given the strange conclusion we came to in the previous section when relying on the geometric definition (\ref{MS}) for the Misner-Sharp mass, let us go back to the original motivation behind that definition, which is gravitational collapse, and examine the latter in the conformally transformed spacetime to find out how the geometric definition (\ref{MS}) is affected.

The main setup needed to study gravitational collapse is, on one hand, a general spherically symmetric metric of the form,
\begin{equation}\label{GeneralSphericalMetric}
\mathrm{d}s^{2}=-e^{2\Theta(t,r)}\mathrm{d}t^{2}+e^{2\Lambda(t,r)}\mathrm{d}r^{2}+R^{2}(t,r)\mathrm{d}o^{2},
\end{equation}
with arbitrary functions $\Theta(t,r)$, $\Lambda(t,r)$ and $R(t,r)$, and on the other hand, an energy-momentum tensor $T^{\mu\nu}=(\rho+p)u^{\mu}u^{\nu}+pg^{\mu\nu}$ associated to matter with mass density $\rho(t,r)$, pressure $p(t,r)$, and 4-velocity $u^{\mu}=(e^{-\Theta},0,0,0)$. The conservation equations $\nabla_{\nu}T_{\mu}^{\,\nu}=0$, then give the following two conditions for the conservation of energy and the conservation of momentum, respectively:
\begin{align}\label{Energy-Momentum Conservation}
\nabla_{\nu}T_{t}^{\,\nu}=0\qquad\mathrm{and}\qquad\nabla_{\mu}T_{r}^{\,\mu}=0.
\end{align}
On the other hand, since $T_{t}^{t}=-\rho$, $T_{r}^{r}=T_{\theta}^{\theta}=T_{\varphi}^{\varphi}=p$ and $T_{t}^{r}=0$, all we are going to need for our purposes here are the $tt$-component, the $rr$-component and the $tr$-component of the field equations. These read, respectively,
\begin{equation}\label{Components}
-\kappa\rho=G_{t}^{\,t}\qquad\mathrm{and}\qquad0=G_{t}^{\,r}.
\end{equation}

Now the original definition introduced by Misner and Sharp in Ref.~\cite{MS} for the mass $m(t,r)$, assumed to be contained inside a spherically symmetric shell of radius $R(t,r)$, is $\partial_{r}m=4\pi R^{2}\rho\partial_{r}R$. This definition follows actually from the very intuitive requirement that within a spherical layer of infinitesimal thickness $\mathrm{d}R$, one finds the element of mass $\mathrm{d}m=4\pi R^{2}\rho\mathrm{d}R$. Therefore, the total mass $m(t,r)$ would simply be,
\begin{equation}\label{RealMS}
m(t,r)=\int4\pi R^{2}\rho\frac{\partial R}{\partial r}\mathrm{d}r.
\end{equation}
It turns out that this integral can straightforwardly be performed after using the set of equations in (\ref{Components}). This comes about thanks to the fact that after multiplying both sides of the first equation in (\ref{Components}) by $R^{2}\partial_{r}R$ and then using the second equation, the right-hand side of the first equation becomes a pure $r$-derivative, so that the integrand in (\ref{RealMS}) becomes an exact differential. The details of the calculations can be found in Ref.~\cite{Urzhumov}. The explicit expression one finds, is
\begin{align}\label{MSFormula}
m(t,r)=\frac{R}{2G}\left(1+e^{-2\Theta}\dot{R}^{2}-e^{-2\Lambda}R'^{2}\right).
\end{align}
Here we have extracted Newton's constant $G$ from $\kappa$. This last result is identical in spherically symmetric spacetimes to the familiar definition (\ref{MS}) often used in the literature. For the sake of completeness, we include here the other important equation in relativistic hydrodynamics obtained by Misner and Sharp in Ref.~\cite{MS}, namely,
\begin{equation}\label{partialtm}
\dot{m}=-4\pi R^{2}p\dot{R}.
\end{equation}
This dynamical equation might be interpreted as relating the variation of the mass $m$ to the power supplied by the pressure $p$ through the spherical screen of radius $R$. This equation is usually written in the form, $D_{t}m=-4\pi R^{2}pD_{t}R$, where the new derivative $D_{t}\equiv e^{-\Theta}\partial_{t}$ has been introduced in Ref.~\cite{MS}. Equation (\ref{partialtm}) is obtained by differentiating (\ref{RealMS}) with respect to time and then using the two conservation equations (\ref{Energy-Momentum Conservation}).

Let us now examine how all these equations, and in particular (\ref{MSFormula}), will be transformed when going to the conformal frame. Starting from the conservation of energy and momentum, the two equations in (\ref{Energy-Momentum Conservation}) become, respectively, after using the fact that in the conformal frame one has $\tilde{\nabla}_{\mu}\tilde{T}_{\nu}^{\,\mu}=-\tilde{T}\partial_{\nu}\ln\Omega$ \cite{Dabrowski} where $\tilde{T}=3\tilde{p}-\tilde{\rho}$ is the trace of the energy-momentum tensor in the conformal frame,
\begin{align}\label{ConfEnergyCons}
\dot{\tilde{\rho}}+(\tilde{\rho}+\tilde{p})\left(\dot{\tilde{\Lambda}}+\frac{2\dot{\tilde{R}}}{\tilde{R}}\right)&=(3\tilde{p}-\tilde{\rho})\frac{\dot{\Omega}}{\Omega},
\nonumber
\\\tilde{p}'+(\tilde{\rho}+\tilde{p})\tilde{\Theta}'&=(\tilde{\rho}-3\tilde{p})\frac{\Omega'}{\Omega}.
\end{align}

The conformally transformed spherically symmetric metric (\ref{GeneralSphericalMetric}) becomes,
\begin{align}\label{ConfSch}
\mathrm{d}s^{2}&=-e^{2\tilde{\Theta}(t,r)}\mathrm{d}t^{2}+e^{2\tilde{\Lambda}(t,r)}\mathrm{d}r^{2}+\tilde{R}^{2}(t,r)\mathrm{d}o^{2}.
\end{align}
Here we have set $e^{\tilde{\Theta}}=\Omega e^{\Theta}$, $e^{\tilde{\Lambda}}=\Omega e^{\Lambda}$ and $\tilde{R}=\Omega R$.
The conformally transformed version of the two field equations in (\ref{Components}), based on the general scheme (\ref{ConfFieldEq}), are, respectively,
\begin{align}\label{ConfComponents}
-\kappa\tilde{\rho}=\Omega^{-2}\tilde{G}_{t}^{\,t}-\Omega^{-4}T^{\Omega t}_{\;\,t}\qquad\mathrm{and}\qquad\tilde{G}_{t}^{\,r}=\Omega^{-2}T^{\Omega r}_{\;\,t}.
\end{align}
Here we have used the fact that $\tilde{\rho}=\Omega^{-4}\rho$ and $\tilde{p}=\Omega^{-4}p$. The definition (\ref{RealMS}) then becomes in the conformal frame
\begin{align}\label{ConfRealMS}
\tilde{m}(t,r)&=\int4\pi\tilde{R}^{2}\tilde{\rho}\frac{\partial\tilde{R}}{\partial r}\mathrm{d}r.
\end{align}
In order to check if (\ref{ConfRealMS}) is integrable, one needs only use the $tt$-component of the conformally transformed field equations (\ref{ConfComponents}) after multiplying both sides of the equation by $\tilde{R}^{2}\partial_{r}\tilde{R}$. A straightforward verification shows that, in contrast to the case we had in the original frame, the resulting term $\tilde{R}^{2}\tilde{R}'\left(\Omega^{-2}\tilde{G}_{t}^{t}-\Omega^{-4}T^{\Omega t}_{\;\,t}\right)$ cannot be written as a pure $r$-derivative and the new Misner-Sharp mass $\tilde{m}(t,r)$ cannot be written in the form (\ref{MSFormula}) by simply putting tildes on the functions $R$, $\Theta$ and $\Lambda$. In other words, when going to the conformal frame the geometric expression $\frac{1}{2G}\tilde{R}\left(1-\tilde{g}^{\mu\nu}\partial_{\mu}\tilde{R}\partial_{\nu}\tilde{R}\right)$ for the conformally transformed Misner-Sharp mass does not hold since this geometric expression would just gives rise to (\ref{MSFormula}) with tildes. In fact, instead of recovering an expression of the form (\ref{MSFormula}) but written with tildes over the letters, one finds the following much more complicated formula:
\begin{align}
\tilde{m}(t,r)&=\frac{\tilde{R}}{2G\Omega^{2}}\left(1+e^{-2\tilde{\Theta}}\dot{\tilde{R}}^{2}-e^{-2\tilde{\Lambda}}\tilde{R}'^{2}\right)
\nonumber
\\&-\int\frac{\tilde{R}^{2}}{2G\Omega^{5}}\left[\frac{2\Omega^{2}\Omega'}{\tilde{R}}-e^{-2\tilde{\Theta}}\left(J\dot{\tilde{R}}+K\tilde{R}'\right)
-e^{-2\tilde{\Lambda}}L\tilde{R}'\right]\!\mathrm{d}r
\end{align}
with
\begin{align}
J&=\frac{2\dot{\Omega}\Omega'}{\Omega}-\dot{\Omega}'+\tilde{\Theta}'\dot{\Omega}+\dot{\tilde{\Lambda}}\Omega'
-\frac{2\dot{\tilde{R}}\Omega'\Omega^{2}}{\tilde{R}},\nonumber
\\K&=\frac{3\dot{\Omega}^{2}}{\Omega}+2\dot{\tilde{\Lambda}}\dot{\Omega}+\frac{4\dot{\tilde{R}}\dot{\Omega}}{\tilde{R}},\nonumber
\\L&=\frac{\Omega'^{2}}{\Omega}+2\tilde{\Lambda}'\Omega'-2\Omega''-\frac{2\tilde{R}'\Omega'(2-\Omega^{2})}{\tilde{R}}.
\nonumber
\end{align}

A relation between the new mass $\tilde{m}(t,r)$ and the original mass $m(t,r)$ can actually be found by expressing the integrand in (\ref{ConfRealMS}) in terms of the areal radius $R$ and the density $\rho$ of the original frame. Indeed, using $\tilde{R}=\Omega R$ and $\tilde{\rho}=\Omega^{-4}\rho$, we have
\begin{align}\label{RealConfMS}
\tilde{m}(t,r)=\int4\pi R^{2}\rho\left(\frac{R'}{\Omega}+\frac{R\Omega'}{\Omega^{2}}\right)\mathrm{d}r.
\end{align}

After performing an integration by parts on the first term and then using (\ref{RealMS}), we obtain
\begin{align}\label{IntByParts}
\tilde{m}(t,r)=\frac{m(t,r)}{\Omega}+\int\left[m(t,r)+4\pi R^{3}\rho\right]\frac{\Omega'}{\Omega^{2}}\mathrm{d}r.
\end{align}

Thus, we recover in the first term the familiar form of the conformal transformation of masses. The last two terms might conveniently be thought of as coming from the 'work' done to 'compress' or 'dilate' the mass. Indeed, the form of the integral suggests the following simple and handy way of looking at the result. The first term could be thought of as representing a contribution from the 'work' $\mathrm{d}W=-m(t,r)\nabla_{r}(1/\Omega)\mathrm{d}r$ done on the matter, whose total mass at the point $r$ is $m(t,r)$, to compress or dilate it along with the spacetime medium during the conformal transformation. The second term, might thereby be thought of as a correction to be brought to the previous term given that the mass $m(t,r)$ itself is conditioned at each point $r$ by the density $\rho(t,r)$.

Before we proceed further with the discussion about the result (\ref{IntByParts}), we would like to make notice here of the fact that, in contrast to what one finds when using the purely geometric definition (\ref{MS}) for the Misner-Sharp mass, one does not obtain the induced energy density as it was the case in (\ref{ConfMSinGR}). One finds, instead, an 'interaction' or 'coupling' term between the original mass $m(t,r)$ and the gradient of the conformal factor $\Omega$, a coupling that was already present, albeit in much more complicated form, in (\ref{ConfSchwMS}).

Now with the integral formula (\ref{RealConfMS}), or its equivalent (\ref{IntByParts}), one will always recover the right conformal factor, as can easily be checked, either for the case of a black hole with a point-like distribution of its interior mass or
for the case of a homogenous FLRW Universe. Indeed, for the Schwarzschild black hole (\ref{SchwMetric}), transformed into (\ref{ConfSchwMetric}), formula (\ref{RealConfMS}) gives, after using $\rho(t,r)=(4\pi r^{2})^{-1}M\delta(r)$ where $\delta(r)$ is the Dirac delta function, given the assumption that the constant mass $M$ of the black hole is concentrated at the center $r=0$ of the latter, the desired result  for the conformal mass: $\tilde{m}(t,r)=M/\Omega$. For the case of the FLRW metric (\ref{FLRW}), transformed into (\ref{ConfFLRW}), formula (\ref{IntByParts}) also gives the desired result, $\tilde{m}(t,r)=m(t,r)/\Omega$, given that in order to preserve the homogeneity of the FLRW Universe, the conformal factor $\Omega(t)$ is allowed to depend only on time.

For a non-uniform mass density $\rho(t,r)$ and/or conformal factor $\Omega(t,r)$ one will still recover for $\tilde{m}(t,r)$ the right conformal factor $\Omega^{-1}$ in front of the total mass $m(t,r)$ but the result will be augmented with the additional integral on the right-hand side of (\ref{IntByParts}). This does not however mean that in this case the Misner-Sharp mass will display a different conformal behavior from that of usual masses. Indeed, even for a usual mass $m$ distributed inside a finite volume $V$, only in the case of uniform densities are we allowed to write $m=\rho V$ from which one recovers the simple relation $\tilde{m}=m/\Omega$. In fact, in the case of the Misner-Sharp mass the additional integral in (\ref{IntByParts}) properly accounts for the effect of a non-uniform conformal deformation of spacetime and/or non-uniform density $\rho(t,r)$. Let us examine the simple case of a non-uniform conformal transformation $\Omega(t,r)$ applied on the Minkowski spacetime $\mathrm{d}s^{2}=-\mathrm{d}t^{2}+\mathrm{d}r^{2}+r^{2}\mathrm{d}o^{2}$ containing a mass density of the form $\rho(t,r)=\lambda/r^{2}$, where $\lambda$ has the dimensions of mass per unit length. The areal radius in this case is then $r$ and the total mass at the coordinate $r$ is $m(t,r)=4\pi\lambda r$. For this case, as we shall see, the simple way of looking at the result (\ref{IntByParts}) suggested above finds a nice illustration. Let us first choose $\Omega(t,r)=(\ell+r)/(2\ell+r)$, where $\ell$ is an arbitrary length\footnote{Note that this specific form of $\Omega$ has been chosen here only for the purpose of simplifying the illustration, as any other form of $\Omega$ that dilates or compresses space would also work.}. The effect of this transformation is to compress space more near the origin at $r=0$ than at infinity where the spacetime is left unchanged. Application of formula (\ref{IntByParts}) then  gives,
\begin{equation}\label{Example1}
\tilde{m}=\frac{m}{\Omega}-\int\frac{8\pi\ell\lambda r}{(\ell+r)^2}\mathrm{d}r=\frac{m}{\Omega}-8\pi\ell\lambda\left[\ln(1+r/\ell)-\frac{r}{\ell+r}\right].
\end{equation}
Here, the constant of integration from the integral has been chosen so that the total mass $\tilde{m}$ vanishes at the origin $r=0$. Let us now choose instead the conformal factor to be $\Omega(t,r)=(\ell+r)/(\frac{1}{2}\ell+r)$. The effect of this transformation is to dilate space more near the origin than at infinity where the spacetime is left unchanged. Application of formula (\ref{IntByParts}) then gives,
\begin{equation}\label{Example2}
\tilde{m}=\frac{m}{\Omega}+\int\frac{4\pi\ell\lambda r}{(\ell+r)^2}\mathrm{d}r=\frac{m}{\Omega}+4\pi\ell\lambda\left[\ln(1+r/\ell)-\frac{r}{\ell+r}\right].
\end{equation}
Here again the constant of integration has been chosen to make the resulting mass $\tilde{m}$ vanish at the origin. Notice that the content of the square brackets in (\ref{Example1}) and (\ref{Example2}) is positive for all $r>0$. Thus, we see that when the conformal transformation acts by compressing space the correction brought to the first term in (\ref{IntByParts}) is negative whereas when the transformation dilates space the correction comes out positive. This justifies the handy way suggested above for interpreting the result (\ref{IntByParts}).

Now let us check how the Misner-Sharp equation (\ref{partialtm}) transforms when going to the conformal frame. For that purpose, let us compute the derivative $\dot{\tilde{m}}(t,r)$ using the integral formula (\ref{RealConfMS}), or better, the result (\ref{IntByParts}). We find, after using the conservation equations (\ref{ConfEnergyCons}) to extract $\dot{\rho}$,
\begin{align}
\!\!\dot{\tilde{m}}(t,r)&=-4\pi\tilde{R}^{2}\tilde{p}\dot{\tilde{R}}\nonumber
\\&+\int\frac{4\pi\tilde{R}^{2}}{\Omega}\left(3\tilde{p}-\tilde{\rho}\right)\left(\tilde{R}'\dot{\Omega}-\dot{\tilde{R}}{\Omega}'\right)\mathrm{d}r\nonumber
\\&+\!\int\frac{4\pi\tilde{R}^{3}}{\Omega^{3}}\left(\tilde{p}+\tilde{\rho}\right)\left(\frac{2\dot{\Omega}\Omega'}{\Omega}
-\dot{\Omega}'+\tilde{\Theta}'\dot{\Omega}+\dot{\tilde{\Lambda}}\Omega'\right)\!\mathrm{d}r.
\end{align}

Finally, let us discuss here how these results are modified when going to scalar-tensor theories of gravity. First, if one takes into account also the energy density associated to the field $\phi$, the total Misner-Sharp mass, to be the analogue of (\ref{RealMS}), should be written as follows:
\begin{equation}\label{TotalMS}
m(t,r)=\int4\pi R^{2}(\rho_{m}+\rho_{\phi})R'\mathrm{d}r,
\end{equation}
where we have denoted by $\rho_{m}$ the energy density $-T_{t}^{t}$ of ordinary matter and by $\rho_{\phi}$ the energy density $-\phi T^{\phi t}_{\,t}$ of the Brans-Dicke field $\phi$. In contrast to equations (\ref{Components}) of general relativity, the proportionality factor between geometry and matter in the $tt$-component, $4\pi(\rho_{m}+\rho_{\phi})=-\frac{1}{2}\phi G_{t}^{\;t}$, of the Brans-Dicke field equations (\ref{BDFieldEq}) is the space-time dependent scalar field $\phi$. Thereby, after multiplying both sides of this equation by $R^{2}\partial_{r}R$, the result will not be a simple $r$-derivative and integral (\ref{TotalMS}) cannot yield such a simple formula as (\ref{MS}). Instead, an integration by parts of (\ref{TotalMS}) gives,
\begin{equation}\label{TotalIntByParts}
m(t,r)=\frac{R\phi}{2}\left(1-g^{\mu\nu}\partial_{\nu}R\partial_{\mu}R\right)
\\-\int\frac{R\phi'}{2}\left(1-g^{\mu\nu}\partial_{\nu}R\partial_{\mu}R\right)\mathrm{d}r.
\end{equation}
We see again that the usual geometric formula (\ref{MS}) is altered in Brans-Dicke theory. Interestingly, however, this formula still yields exactly the same result found for the FLRW metric in (\ref{ConfMSinBD}), given that the conformal factor $\Omega(t)$ is only time-dependent. Notice also that one is still able to introduce here a different definition for the Misner-Sharp mass in scalar-tensor theories of gravity in order to make it yield the geometrical definition (\ref{MS}). Indeed, the new definition need simply read,
\begin{equation}\label{DiffBDMS}
m(t,r)=\int4\pi R^{2}\left[(\rho_{m}+\rho_{\phi})R'+\sigma\phi'\right]\mathrm{d}r,
\end{equation}
with the function $\sigma(t,r)$ satisfying the following partial differential equation:
\begin{equation}
\partial_{r}\sigma=\frac{R'}{\phi}(\rho_{m}+\rho_{\phi})-\frac{2\sigma}{R}.\nonumber
\end{equation}
This definition, however, is only ad hoc and cannot be justified apart from the desire to recover the familiar geometric definition. Moreover, this new definition does not coincide with Nariai's new definition of the Misner-Sharp mass he proposed in Ref.~\cite{Nariai} for scalar-tensor theories of gravity.

As for the conformal transformation of the formula (\ref{TotalMS}), one finds, by virtue of the conformal invariance of the Brans-Dicke field equations, the simple result
\begin{equation}\label{TotalIntByParts}
\tilde{m}(t,r)=\frac{\tilde{R}\tilde{\phi}}{2}\left(1-\tilde{g}^{\mu\nu}\partial_{\nu}\tilde{R}\partial_{\mu}\tilde{R}\right)
\\-\int\frac{\tilde{R}\tilde{\phi}'}{2}\left(1-\tilde{g}^{\mu\nu}\partial_{\nu}\tilde{R}\partial_{\mu}\tilde{R}\right)\mathrm{d}r.
\end{equation}
This result reproduces again the formula (\ref{MSinBD}), found using the definition (\ref{MS}) because the conformal factor there does not depend on $r$. For more general functions of $\Omega$, though, it is not possible to recover formula (\ref{MSinBD}).
\section{Conclusion}
The way the Misner-Sharp mass is transformed under conformal mappings of spacetime has been examined in detail. We showed that the conformal transformation of the mass is different whether one uses the geometric definition of the latter or goes back to its initial material origin as introduced by Misner and Sharp. We found that the mass transforms as usual masses do only when one relies on the material definition. Moreover, we showed that when using the latter, only the effect of the conformal deformation of spacetime on the mass emerges without the appearance of the induced geometric energy density as is the case whenever one starts from the geometric definition. This means that the original definition sees only the matter content as well as that part of geometry that affects directly the matter content, whereas the geometric definition gives the geometric equivalent of the matter content. Therefore, adopting the geometric definition of the Misner-sharp mass when performing a conformal transformation of spacetime will necessarily not give the true conformal transformation of the matter content but only its geometric equivalent, whence the different conformal factors the two definitions yield.

It must be mentioned here that besides the Misner-Sharp mass, the so call Hawking-Hayward mass \cite{Hawking, Hayward2}, is also found to transform in the same way as the geometric definition of the former does, i.e. it acquires the 'wrong' factor \cite{Prain}. Given that the original Hawking-Hayward mass is of a purely geometric nature, it is no wonder that the latter would transform in a different way from the former. However, given that the Hawking-Hayward mass gives back the mass of the material content of the region of spacetime under consideration, a conformal transformation of the mass would certainly be different whether one expresses the geometry first in terms of its matter content through the field equations before performing the conformal transformation, or applies the conformal transformation directly on the geometric definition. A detailed study of this issue will be attempted in a separate work.
\\\\
\textbf{Acknowledgement}:
I would like to thank the anonymous referee for his/her valuable and helpful comments.

\end{document}